\documentclass[preprint]{aastex}
\usepackage{psfig}
\def\HI {H\kern0.1em{\sc i}} 
\def\etal   {{\sl et~al.\ }}

\begin{document}
\title{~~\\ ~~\\ Kinematic Age Estimates for 4 Compact Symmetric
  Objects \\ from  the Pearson-Readhead Survey}
\author{G. B. Taylor\altaffilmark{1}, J. M. Marr\altaffilmark{2}, 
T. J. Pearson\altaffilmark{3}, \& 
A. C. S. Readhead\altaffilmark{3}}

\altaffiltext{1}{National Radio Astronomy Observatory, Socorro, NM
  87801, USA}
\email{gtaylor@nrao.edu}
\altaffiltext{2}{Union College, Schenectady, NY 19308, USA }
\email{marrj@union.edu}
\altaffiltext{3}{California Institute of Technology, Pasadena, CA
  91125, USA }
\email{tjp@astro.caltech.edu; acr@astro.caltech.edu}

\received{2000 March 24}
\accepted{2000 May 9}

\slugcomment{Accepted to the Astrophysical Journal}

\begin{abstract}

  Based on multi-epoch observations at 15 and 43 GHz with the Very
  Long Baseline Array (VLBA) we detect significant angular expansions
  between the two hot spots of 4 Compact Symmetric Objects (CSOs).
  From these relative motions we derive kinematic ages of between 300
  and 1200 years for the radio emission.  These ages lend support to
  the idea that CSOs are produced in a recent phase of activity.
  These observations also allow us to study the evolution of the hot
  spots dynamically in individual sources.  In all 4 sources the hot
  spots are separating along the source axis, but in 1031+567 the tip
  of one hot spot appears to be moving almost orthogonally to the
  source axis.  Jet components, seen in 3 of the 4 sources observed,
  are found to be moving relativistically outward from the central
  engines towards the more slowly moving hot spots.

\end{abstract}


\section{Introduction}

While most radio galaxies exhibit an asymmetric, core-jet morphology
on the parsec scale,
approximately 7\% of radio galaxies in complete,
flux-limited samples display parsec-scale jets and lobe emission 
on both sides of a central engine (Taylor \etal\ 1996).
This emission is thought to be free of Doppler boosting effects
(Wilkinson \etal\ 1994) and the sources are believed to be
physically small as opposed to appearing small due to projection. 


The measurement of a kinematic age for the CSO 0710+439 by Owsianik \&
Conway (1998) of just 1100 $\pm$ 100 years lent strong support to the
theory that CSOs are small by virtue of their youth and not because of
confinement.  This was the favored interpretation by Phillips \& Mutel
(1980, 1982) who first drew attention to a group of 4 compact double
sources with steep spectra and slow motions compared to the majority
of core-jet sources.  Although further multi-frequency VLBI
observations have revealed that 2 of these (CTD93 - Shaffer, Kellermann \& 
Cornwell 1999; and 3C\,395 - Taylor 2000) are actually asymmetric core-jet
sources, many of the speculations of Phillips \& Mutel have been borne
out.  Their misidentification of two sources emphasizes the fact that
sensitive multi-frequency VLBI observations are required to
demonstrate symmetric structure on the parsec scale.  In particular,
not all (or even most) GHz Peaked Spectrum (GPS) radio galaxies belong
to the CSO class.  Failure to find any evidence of the extremely dense
environment required to confine CSOs, along with some indirect age
measurements, also led Readhead \etal\ (1996a) to favor the idea that
CSOs are young.  In fact, angular separation rates for the CSOs
measured to date indicate typical hotspot velocities of $\sim$0.1
$h^{-1}$ c ({\it e.g.,} Owsianik et al.\ 1998, Owsianik \& Conway 1998,
Polatidis et al. 1999).  Such speeds are about an order of magnitude
larger than predicted by Readhead \etal\ (1996a) and so indicate even
younger ages.

The evolution of CSOs is of considerable current interest.  To
understand the evolution of radio galaxies we need to be able to
recognize the evolutionary state of any given galaxy.  Readhead \etal
(1996b) have proposed an evolutionary model for powerful radio
sources in which CSO's evolve first into Compact Steep Spectrum
doubles and then into large Fanaroff-Riley (1974) Type II objects.
Based on a different sample of somewhat larger objects, Fanti \etal
(1995) reached the same conclusions.  Measurements of source 
ages are crucially important to the understanding of the evolution of
this intriguing class of objects. 

In 1994 Taylor, Readhead, and Pearson (1996) performed VLBA
observations of 4 CSOs from the Pearson-Readhead survey (Pearson \&
Readhead 1988) at 15 GHz in order to pinpoint their centers of
activity.  The results were surprising in that the components inferred
to be the cores, because of strongly inverted spectra and compactness,
were found not to be associated with the strongest central components
seen at 5 GHz and lower frequencies, but instead were 
unresolved components very close to the midway point between the
hotspots ({\it e.g.,} Fig.~1).  The bright components seen at 5 GHz (in some
cases previously mis-identified as cores) turn out to be one-sided
jets.  To further characterize the properties of the core components
we carried out 43 GHz VLBA observations of several PR CSOs in 1996
and 1999.

In \S3 we present multi-epoch VLBA observations at 15 and 43 GHz 
for four CSOs from the PR survey.   These results are used in \S4
to measure the advance speeds of the hot spots, and to thereby
determine kinematic age estimates.  Velocity measurements are 
expressed in terms of $h$ = H$_0$ / 100 km s$^{-1}$ Mpc$^{-1}$,
and where physical scales are quoted (or drawn) we assume $h = 0.65$.

\section{Observations and Data Reduction}

Observations were made at 15 and 43 GHz using the Very Long
Baseline Array (VLBA)\footnote{The National Radio Astronomy
Observatory is a facility of the National Science Foundation
operated under cooperative agreement by Associated Universities, Inc.} 
telescope
at multiple epochs between 1994 and 1999.  The details of the
observations are provided in Table 1.  All sources were observed with
5--8 scans spread across a wide range in hour angle in order to obtain
good ($u$, $v$) coverage.  The VLBA correlator produced typically 16
frequency channels across every 8 MHz of observing bandwidth in each 2
second integration period.

Calibration procedures were followed for the 15 GHz data in a manner
similar to that used for the first epoch observations as described in
Taylor, Readhead \& Pearson (1996).  At 43 GHz the delays and IF phase
offsets were determined from the injected pulse-calibration and a
short observation of the strong calibrator 3C\,84.  The data were then
averaged in frequency to a single channel of 64 MHz.  No global
fringe-fitting was attempted as the scatter in the solutions so
derived was judged to be worse than the small residual delay error.
After phase self-calibration with a 10 s solution interval and a
point-source model, the data were coherently averaged to 10 s
integrations.  All editing, imaging, deconvolution, and
self-calibration were performed using {\sc Difmap} (Shepherd, Pearson
\& Taylor 1994, 1995).  Several iterations of phase self-calibration
and imaging were performed with each data set before any attempt at
amplitude self-calibration, and no amplitude self-calibration was
attempted at 43 GHz.  No reliable 43 GHz image could be obtained for
the weakest source, 0108+388. At each iteration, windows for clean
components were added, if necessary, to provide support and reject
sidelobes.

Once the data were completely self-calibrated, Gaussian model-fitting
was performed using {\sc Difmap}.  To determine relative motions
the component shapes and sizes were frozen in the model-fitting to equal
the fitted values in the 
first epoch, while the flux densities and positions of components were
allowed to vary.  The AIPS task JMFIT was also used to perform 
Gaussian model-fitting to the hotspots in the image plane.
In all cases the AIPS JMFIT produced very similar results to fitting the 
visibility data.


\section{Results}

A history of the VLBI observations of each source studied herein can
be found in Taylor, Readhead \& Pearson (1996; hereafter TRP96).  For
convenience source identifications and redshifts are given in Table 2,
and measurements of core properties are summarized in Table 3. Here we
describe our velocity measurements and compare them with recent
results in the literature.  We compute the kinematic age of the source
in its rest frame, $\tau_k$, as $\tau_k = \theta_{\rm hs}/\mu_{\rm hs}
( 1+z)$ where $\theta_{\rm hs}$ is the angular separation of the hot
spots, and $\mu_{\rm hs}$ is the angular separation rate of the hot
spots.

\subsection{0108+388}

In Fig.~1 we show a naturally weighted 15 GHz image of 0108+388.  This
image shows continuous emission connecting the two main components
(labeled C1 through C7).  Component C3 was identified as the core
based on its inverted spectrum and compactness (TRP96; 
and see Table 3).
A detailed spectral analysis between 1.6 and 15 GHz has been carried
out by Marr, Taylor, and Crawford (2000a,b) who find that while  
C3 is likely to have
a spectral turnover at a high frequency due to synchrotron
self-absorption, the spectra of other components turn over around
5 GHz because of free-free absorption 
by an ionized disk centered on the nucleus.  More evidence for a
relatively dense circumnuclear environment comes from \HI\ absorption
measurements by Carilli et al.\ (1998) who find an optical depth of 
0.44 $\pm$ 0.04 and implied column density of 
80.7 $\times$ 10$^{18}$ T$_s/f$ cm$^{-2}$, where T$_s$ is the spin temperature 
and $f$ is the \HI\ covering factor.  Owsianik, Conway \& Polatidis (1998)
measured an angular separation rate for the outer components (C1 and
C7) of 9.3 $\pm$ 1.2 $\mu$as yr$^{-1}$ from 3 epochs of
global VLBI observations at 5 GHz spread over 12 years.


Given the identification of C3 as the core in 0108+388, the relative velocity
of 0.57 $\pm$ 0.35 $h^{-1}$c (see Fig.~1 and Table 4) is at first
glance somewhat alarming since one traditionally assumes the core
component to be stationary.  The core component, however, is likely to
be just the optically thick base of the jet (Blandford \& K\"onigl
1979), and could well appear to move in either direction along the jet
axis as new jet components emerge.  In one of the few cases where
absolute motions have been obtained there is evidence that the
``core'' of 1928+738 moves in this way (Ros \etal\ 1999).  For this
reason, the core component makes a poor choice as a reference feature.
Instead we adopt the westernmost component (C1) as the reference
feature.  As we will argue, this is likely to be a subrelativistic
hotspot.  This choice is somewhat arbitrary, and we could just
as well have taken the easternmost hot spot (C7) as a reference. 

From our 3 VLBA epochs at 15 GHz we find a separation rate for C1 and
C7 of 11 $\pm$ 2 $\mu$as yr$^{-1}$, which is in agreement with
the measurements of Owsianik \etal (1998) and also with recent
measurements at 8.4 GHz (Polatidis \etal\ 1999).
Our kinematic age estimate for 0108+388 is 310 $\pm$ 70 yrs.
Assuming that C1 and C7 are moving apart at equal speeds, this gives
each an advance speed of 0.12 $h^{-1}$c.  The best-fit models from which
these velocities were derived are listed in Table 4.  A significant
velocity of 0.79 $\pm$ 0.04 $h^{-1}$c is found for component C5, although since
this is relative to C1 the true velocity of C5 likely to be
smaller by 0.12 $h^{-1}$c. 



Due to the low flux density of 0108+388 at 43 GHz, it was not possible
to self-calibrate the data and make an image.  Even so we can still
estimate that the core flux density must be less than $\sim$40 mJy
at 43 GHz,
otherwise it would have been readily detected.  This indicates that
the spectral index is not as steeply inverted as at the lower 
frequencies (see Table 3).


\subsection{0710+439}

Owsianik \& Conway (1998) reported the first significant detection of
a hot spot advance speed based on 5 epochs on 0710+439 at 5 GHz
between 1980 and 1993.  The hot spots (components A2 and C2)
were found to have a separation rate of 14 $\pm$ 1.6 $\mu$as
yr$^{-1}$.


In Fig.~2 we show our 2nd epoch 15 GHz observations with motions
derived from the two epochs indicated by arrows.  Components A and C
are both leading-edge brightened with emission fading gradually
towards the center of the source.  Component B is more compact at the
southern end and becomes wider to the north with an opening angle of
$\sim$20 degrees.  At the base of the jet a compact inverted spectrum
component (B5 in Fig.~2)
was identified and inferred to be the core by TRP96.
Gaussian model-fits and component motions are given in Table 5.
In general our 15 GHz model components correspond to the 5 GHz 
model of Owsianik \& Conway (1998), but at 15 GHz the extended 
components A1, B1 and C1 are resolved out.  The other difference 
in the 5 and 15 GHz models is the core component (B5) which was 
too weak to be included in the 5 GHz model.  


We find a separation rate between the hotspots (A2 and C2) of 29 $\pm$
8.7 $\mu$as yr$^{-1}$.  The implied velocity of advance, assuming
equal speeds, is 0.26 $h^{-1}$c.  This velocity is nearly twice that
found by Owsianik \& Conway (1998).  The kinematic age we derive
for 0710+439 is 550 $\pm$ 160 yrs.
Since the epochs do not overlap,
one explanation could be that the hot spot advance speed has recently
doubled.  A more likely explanation for the discrepancy is that the 15
GHz observations are more sensitive to the motion of a bright, compact
working surface, while the 5 GHz observations measure the more 
stable overall
expansion of the lobe.  In support of this idea we note
that the size of component A2 is 0.51 $\times$ 0.33 mas at 15 GHz, and
0.84 $\times$ 0.57 mas at 5 GHz.  


We also find a substantial velocity of 1.36 $\pm$ 0.16 $h^{-1}$c for
component B3 relative to C2.  This velocity, however, is accompanied
by a change in the flux density of B3 from 116 mJy in 1994.971 to only
63 mJy in 1999.587.  The large drop in flux could indicate a region or
subcomponent that faded significantly, causing a large shift in the
centroid.

In Fig.~3 we show an image of 0710+439 at 43 GHz.  Only the core,
inner jet complex, and brightest region of the northern hot spot are
detected.  The hot spot is well resolved in these observations and
shows a similar twist to the southeast as indicated in the 15 GHz
image.  The core component is prominent and still unresolved.  The
core has a 15-43 GHz spectral index of $-$0.2 $\pm$ 0.3 (where $S_\nu
\propto \nu^\alpha$).  Despite the higher resolution afforded by the
43 GHz observations, no significant component motions were detected due
to the shorter time baseline and lower signal-to-noise compared to the
15 GHz observations.

\subsection{1031+567}


A map made from our 15 GHz observations of 1031+567 is shown in Fig.~4 and
the modelfit components are listed in Table 6.
No central core component is apparent.  It is likely that the 
two outermost components are the working surfaces and lobes of two oppositely
directed jets, 
given the edge brightened appearance and steep
spectra (TRP96) and that no core emission is detected.
This source is also remarkable for the very small (1 -- 9\%) change
in the flux density of its components.

We report here a first tentative detection of the expansion 
of 1031+567.  We find a separation rate between the 
hot spots W1 and E1 of 14.6 $\pm$ 4.8 $\mu$as yr$^{-1}$,
although not along the axis of the source (see Fig.~4).  A 
larger expansion rate is found between W1 and E2 of 
37.6 $\pm$ 8.4 $\mu$as yr$^{-1}$ roughly along the position
angle of the axis.  The kinematic age derived from the 
W1--E2 expansion rate is 620 $\pm$ 140 yrs.  Assuming 
equal advance speeds this corresponds to a velocity of 0.31 $h^{-1}$c.

\subsection{2352+495}

The hot spots (labeled A and C after Conway \etal\ 1992) are well
resolved in our 15 GHz image (Fig.~5).  We also see a faint narrow jet
to the south of the bright B1-B5 complex with a compact component (D)
embedded in it and identified by TRP96 as the core.
Based on global VLBI observations at 5 GHz over 14 years, 
Owsianik, Conway \& Polatidis (1999) measured a relative angular
separation rate for the hot spots of 21.1 $\pm$ 2.7 $\mu$as yr$^{-1}$.  


The model-fit results at 15 GHz for 2352+495 are listed in Table 7.
From our two epoch VLBA observations over 4.6 years we find an angular
separation velocity for the hot spots of 33 $\pm$ 11 $\mu$as
yr$^{-1}$.  Given our large uncertainty, these observations are
consistent with the findings of Owsianik \etal\ (1999).  Our measured
component separation rate yields a kinematic age for 2352+495 of 1200
$\pm$ 400 years.  Assuming equal advance speeds for the hot spots
yields a velocity of 0.16 $h^{-1}$c. The jet components in the 
component B complex have velocities ranging from 0.27 -- 0.76 $h^{-1}$c.

In Fig.~6. we show a 43 GHz image of 2352+495.  Only the core and
B1-B5 complex is detected.  The core component appears well isolated
and still unresolved, and has a 15-43 GHz spectral index of $-$0.6
$\pm$ 0.4.

\section{Discussion}

\subsection{Baby Radio Galaxies?}

The kinematic ages derived for the 4 bright CSOs studied here range
from 300 to 1200 yrs.  This is part of a growing body of evidence
conclusively demonstrating that CSOs are indeed young objects.
These ages are derived under the assumption
that the velocity has been constant over the entire period of
activity.  The kpc-scale bridge of emission in 0108+388 (Baum 
\etal\ 1990) suggests prior active phases in at least this
CSO.  Owsianik \etal\ (1998) have suggested that the one-sided
appearance of the large scale structure is due to this 
periodic nature and light travel time effects.


In 3 of the 4 CSOs observed here the hot spots appear to be advancing
along the jet axis.  The one exception is 1031+567 for which the
separation of the hot spot heads is not along the axis of the source,
although the lobe complex is moving out along the source axis.  In the
dentist drill model (Scheuer 1974) for radio galaxy evolution, the
hotspots wander around the leading edge of the lobe.  At lower
frequencies (see for example Owsianik \etal\ 1999) the northern lobe
of 2352+495 appears to extend further from the core than the hotspot.
There is also a sharp bend at the hotspot so it is possible that the
jet is deflected at the ``primary'' hot spot toward the east.


\subsection{Core and Jet Properties}

Three of the four CSO's studied here contain central components
inferred to be the cores.  All three of these apparent core components
display spectra that peak around 15 GHz (see Table 3) and are
unresolved.  If the turnovers are due to synchrotron self-absorption,
following Marscher (1983) one finds that the magnetic fields in the
cores must be less than 7 $\times 10^4$, 80, and 600 Gauss in
0108+388, 0710+439, and 2352+495, respectively.  Free-free absorption
is unlikely to be significant at 15 GHz, although it
has been demonstrated to be significant at frequencies below 5 GHz in
the central regions of some CSOs ({\it e.g.,} 1946+708 -- Peck, Taylor, \&
Conway 1999; 0108+388 --  Marr, Taylor, \& Crawford 2000a,b). 
One might also expect the cores to turn over more sharply than is
indicated in Table 3 if free-free absorption is the dominant 
mechanism.




In 0108+388 we find marginal evidence for motion of the core
component.  Less significant motions are detected in 0710+439 and
2352+495.  These motions are expected in the Blandford \& K\"onigl
(1979) model where the core is the base of an optically thick jet as
new components emerge from the center of activity.

Jet components moving relativistically from the cores towards the hot
spots are seen in all sources studied here except 1031+567.  In
0710+439 and 2352+495 the jet components appear stronger on one side
of the core, consistent with their fast motions and probably
indicating that these sources do not lie in the plane of the sky.
Without some additional constraint on jet velocities, however, the
inclination angle is not well determined.

\section{Conclusions}

We confirm hot spot advance speeds of $\sim$0.2 $h^{-1}$c in three CSOs and
present a new detection for the CSO 1031+567.  
These growth rates correspond to ages between 300 and 1200 yrs
for the current phase of activity in the sources studied here.
Further observations should allow for a more precise determination
of the hot spot velocities and will test the
dentist drill model for radio sources.  


\acknowledgments

We thank the referee, Ken Kellermann, for insightful comments on the
manuscript.  This research has made use of the NASA/IPAC Extragalactic
Database (NED) which is operated by the Jet Propulsion Laboratory,
Caltech, under contract with NASA.  This research has made use of data
from the University of Michigan Radio Astronomy Observatory which is
supported by the NSF and by funds from the University of Michigan.

\clearpage

\begin{figure}
\vspace{19cm}
\includegraphics{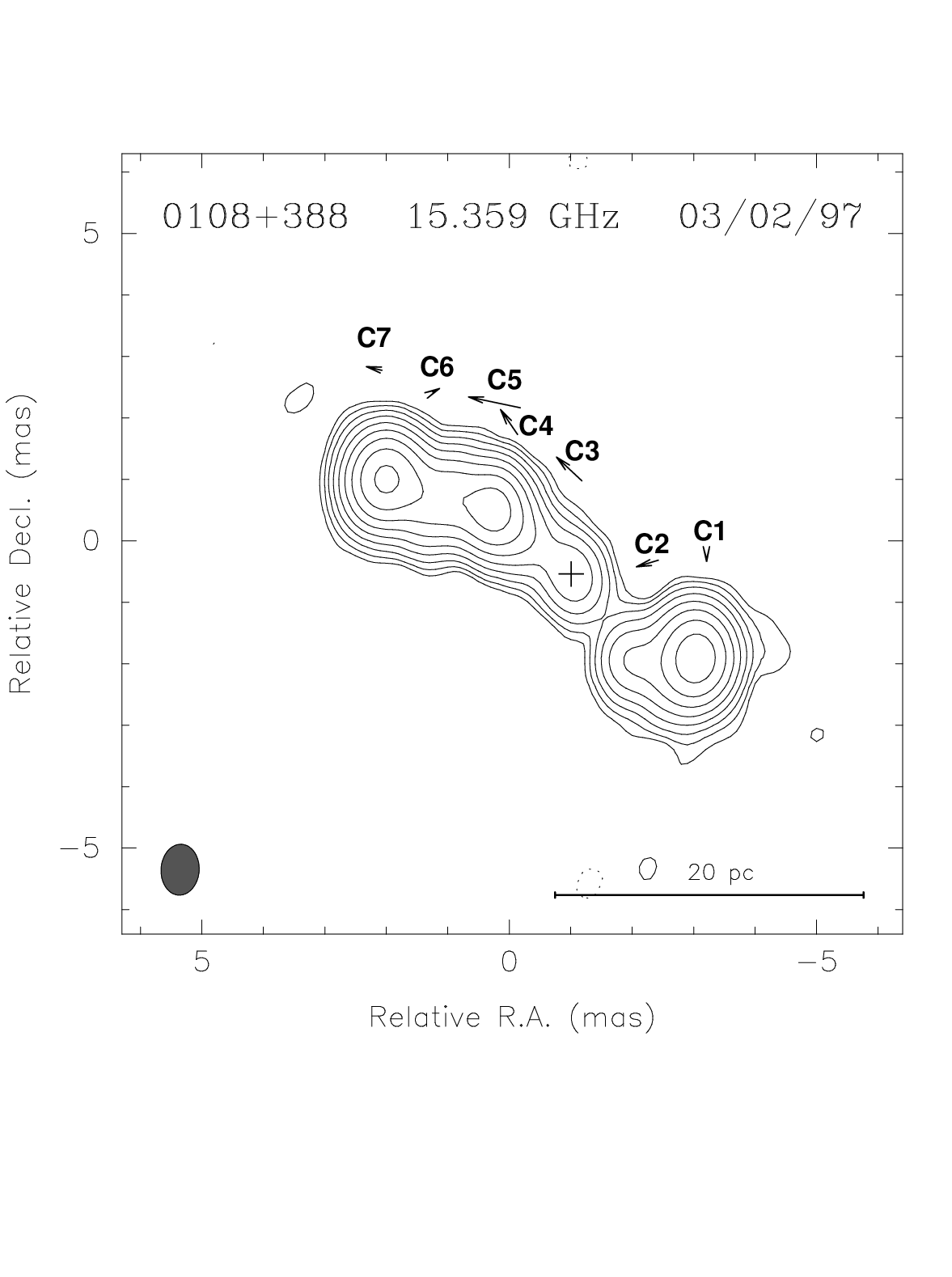}
\caption{The naturally weighted 15 GHz VLBI image of 0108+388
from the February 1997.092 epoch.  
Contours are drawn at $-$0.4, 0.4, 0.8, 1.6, 3.2, ..., 64 mJy/beam. 
The peak in the image is 123 mJy/beam.
The cross
marks the location of the center of activity as discussed in the text.
The synthesized beam FWHM is drawn in the lower left-hand corner and
has dimensions 0.83 $\times$ 0.62 mas in position angle $-$4\arcdeg.
Arrows indicate the direction of component motions, but the lengths have been 
magnified by a factor of 5 for legibility.  
\label{fig1}}
\end{figure}
\clearpage

\begin{figure}
\vspace{19cm}
\includegraphics{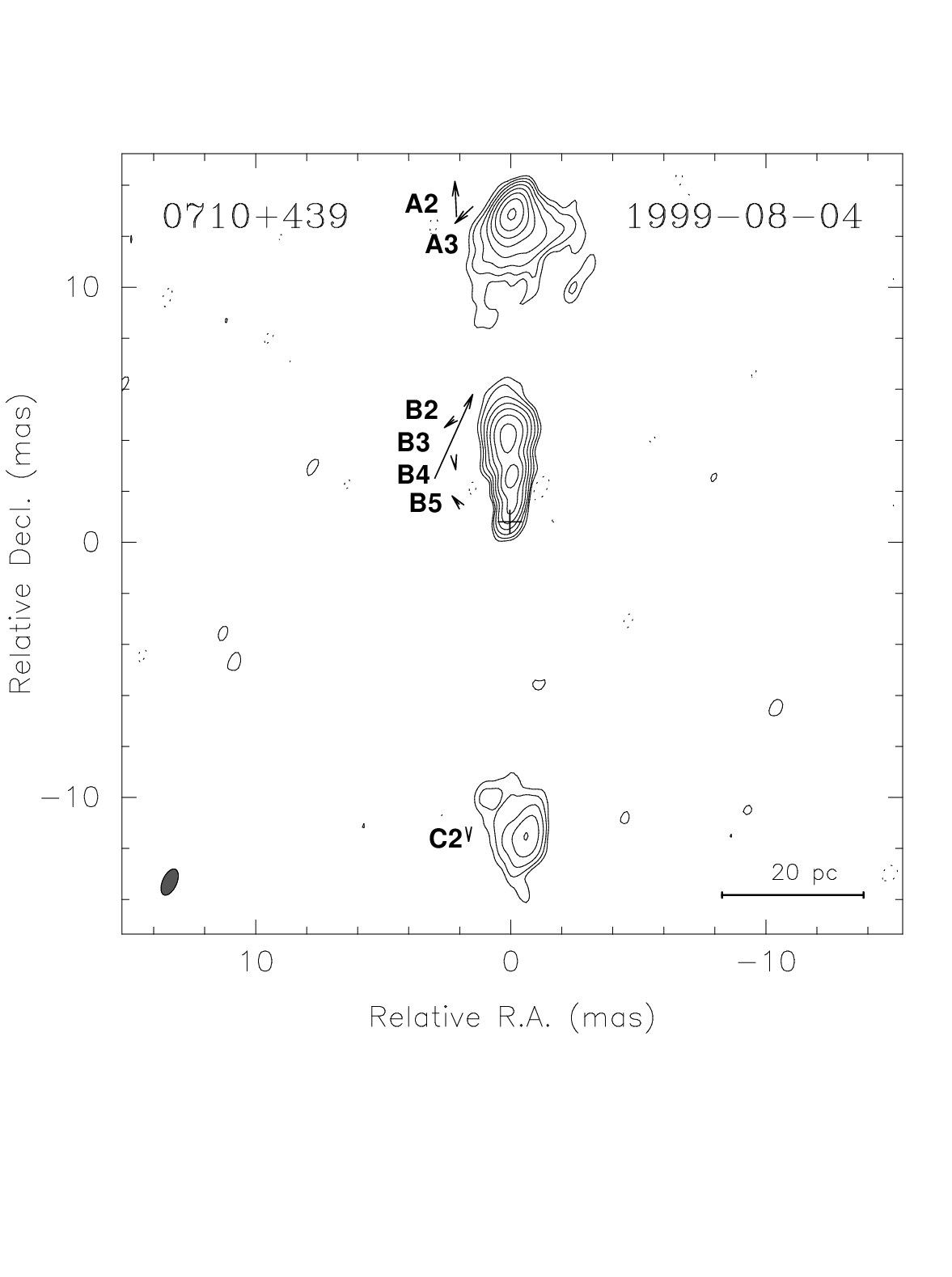}
\caption{The naturally weighted 15 GHz VLBI image of 0710+439 
from the August 1999.587 epoch.  
Contours are drawn at $-$1, 1, 2, 4, ..., 128 mJy/beam. The peak
in the image is 142 mJy/beam. The cross
marks the location of the center of activity as discussed in the text.
The synthesized beam FWHM is drawn in the lower left-hand corner and has 
dimensions 1.09 $\times$ 0.55 mas in position angle $-$24\arcdeg.
Arrows indicate the direction of component motions, but the lengths have been 
magnified by a factor of 10 for legibility.
\label{fig2}}
\end{figure}
\clearpage

\begin{figure}
\vspace{19cm}
\includegraphics{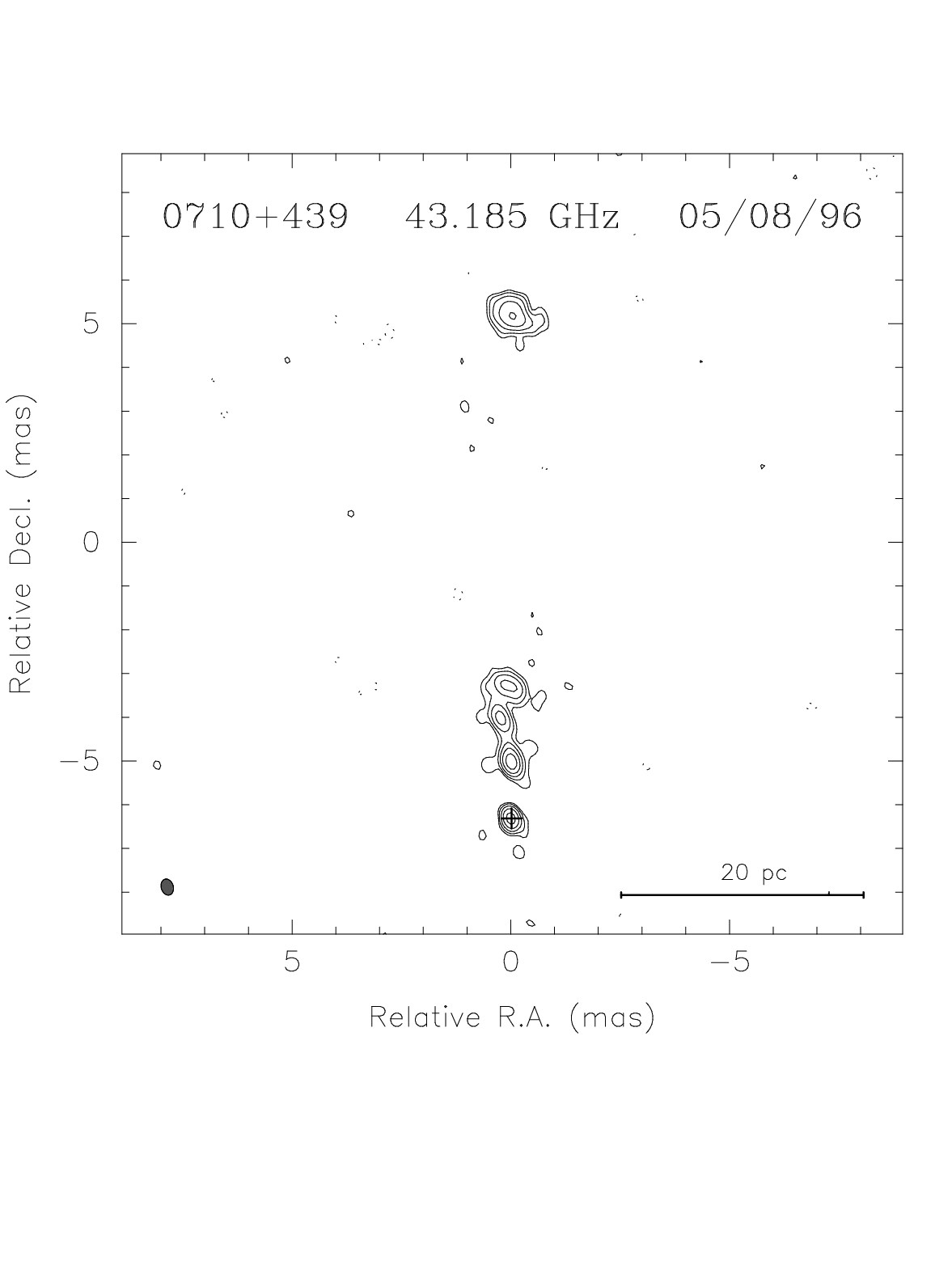}
\caption{The naturally weighted 43 GHz VLBI image of 0710+439 
from the 1996 epoch.  
Contours are drawn at $-$1.5, 1.5, 3, 6, 12 and 24 mJy/beam. The peak
in the image is 39 mJy/beam. The cross
marks the location of the center of activity as discussed in the text.
The synthesized beam FWHM is drawn in the lower left-hand corner and has 
dimensions 0.38 $\times$ 0.28 mas in position angle 17\arcdeg.
\label{fig3}}
\end{figure}
\clearpage

\begin{figure}
\vspace{19cm}
\includegraphics{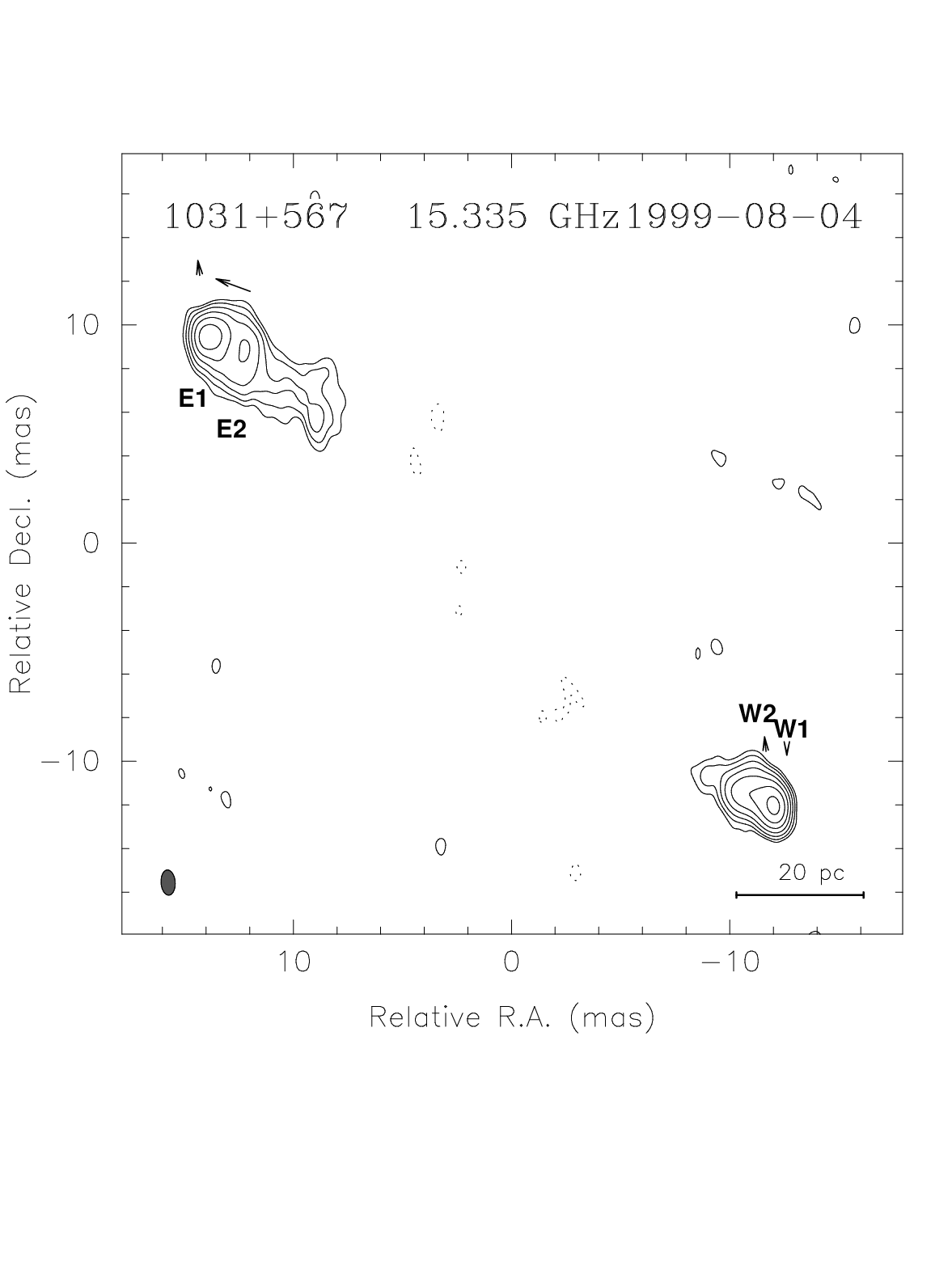}
\caption{The naturally weighted 15 GHz VLBI image of 1031+567
from the 1999.587 epoch.  
Contours are drawn at $-$0.8, 0.8, 1.6, 3.2, ..., 51.2 mJy/beam. The peak
in the image is 66 mJy/beam.
The synthesized beam FWHM is drawn in the lower left-hand corner and has 
dimensions 1.15 $\times$ 0.65 mas in position angle 4\arcdeg.
Arrows indicate the direction of component motions, but the lengths have been 
magnified by a factor of 10 for legibility.
\label{fig4}}
\end{figure}
\clearpage

\begin{figure}
\vspace{19cm}
\includegraphics{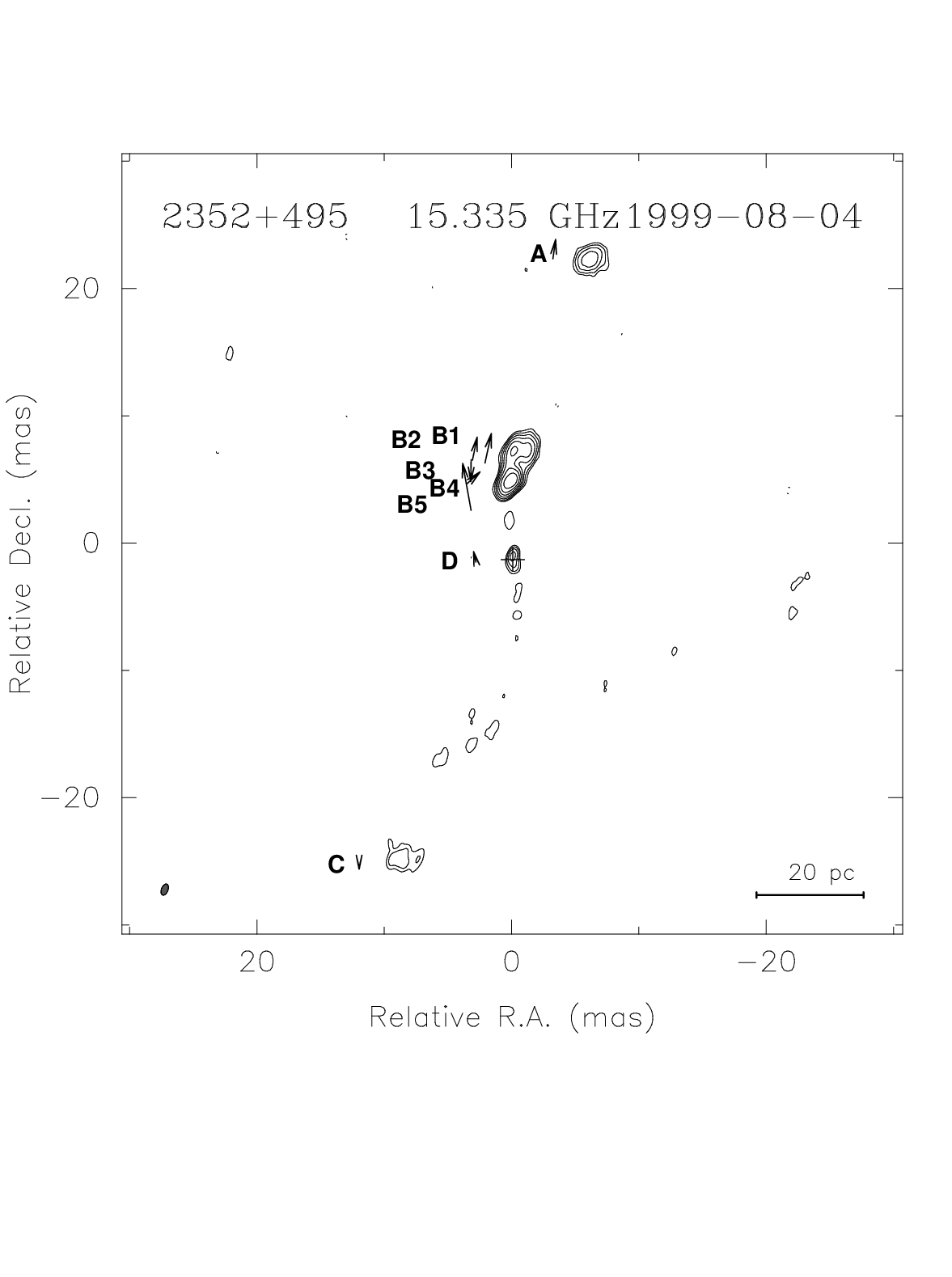}
\caption{The naturally weighted 15 GHz VLBI image of 2352+495 
from the 1999.587 epoch.  
Contours are drawn at $-$1, 1, 2, 4, ..., 64 mJy/beam. The peak
in the image is 95 mJy/beam. The cross
marks the location of the center of activity as discussed in the text.
The synthesized beam FWHM is drawn in the lower left-hand corner and has 
dimensions 0.91 $\times$ 0.51 mas in position angle $-$21\arcdeg.
Arrows indicate the direction of component motions, but the lengths have been 
magnified by a factor of 10 for legibility.
\label{fig5}}
\end{figure}
\clearpage

\begin{figure}
\vspace{19cm}
\includegraphics{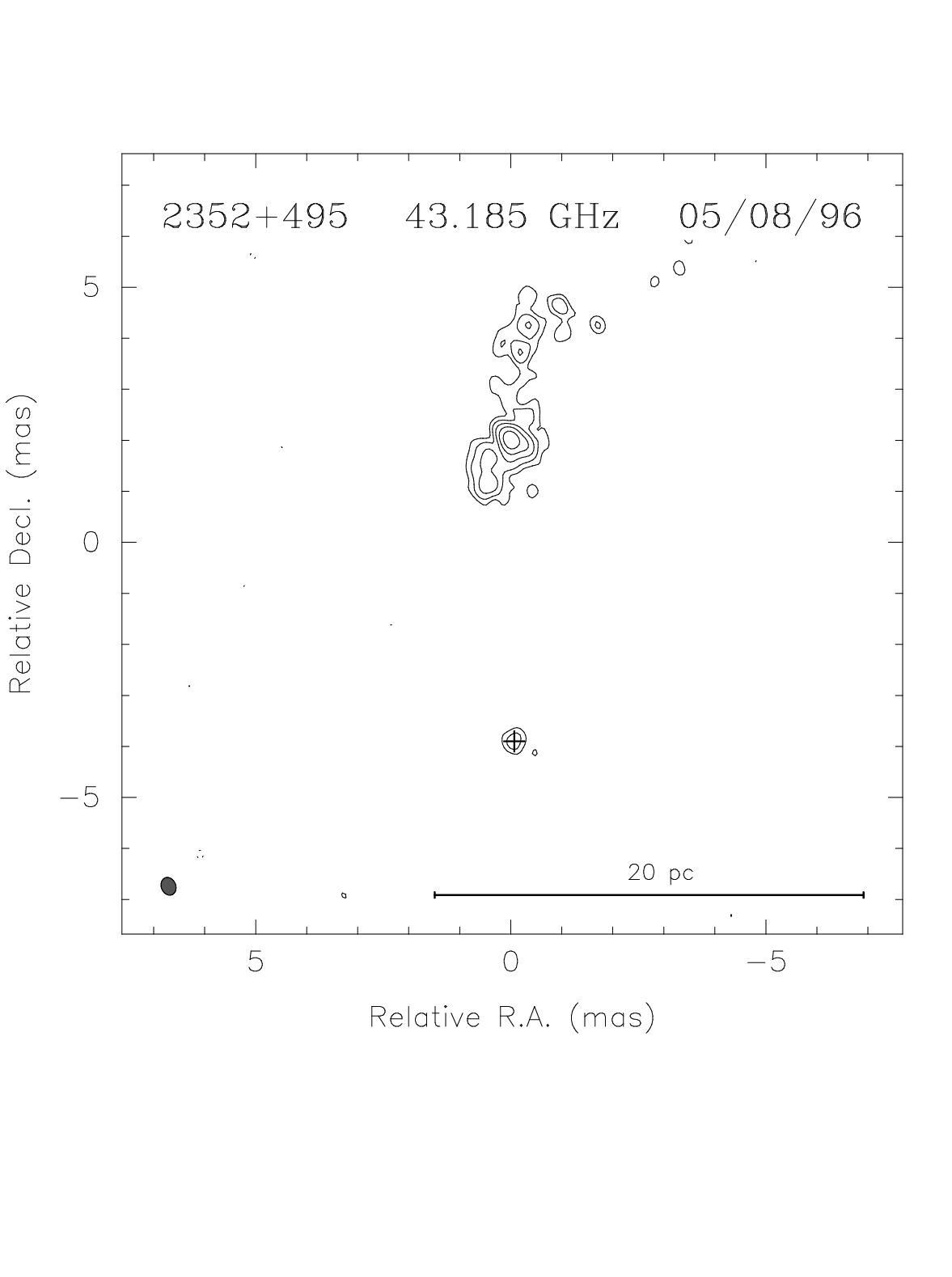}
\caption{The naturally weighted 43 GHz VLBI image of 2352+495
from the August 1996 epoch.  
Contours are drawn at $-$1.5, 1.5, 3, 6, 12 and 24 mJy/beam. The peak
in the image is 43 mJy/beam. The cross
marks the location of the center of activity as discussed in the text.
The synthesized beam FWHM is drawn in the lower left-hand corner and has 
dimensions 0.36 $\times$ 0.28 mas in position angle 23\arcdeg.
\label{fig6}}
\end{figure}
\clearpage

\def\dg{$^{\circ}$}
\begin{center}

TABLE 1 \\
\smallskip
O{\sc bservational} P{\sc arameters}
\smallskip

\begin{tabular}{l l r r r r r r r r}
\hline
\hline
Source & Date & Frequency & Bandwidth$^a$ & Scan Length & Total Time \\
 &  & (GHz) & (MHz) & (min) & (hours) \\
\hline
\noalign{\vskip2pt}
0108+388 & 1994.971 & 15.4 & 16 & 26 & 2.2 \\
         & 1996.594 & 43.2 & 64 &  5 & 1.3 \\
         & 1997.092 & 15.4 & 64 &  5 & 4.3 \\
         & 1999.587 & 15.4 & 64 &  5 & 1.2 \\
         & 1999.587 & 43.2 & 64 &  7 & 1.6 \\
0710+439 & 1994.971 & 15.4 & 16 & 13 & 1.3 \\
         & 1996.594 & 43.2 & 64 &  5 & 1.3 \\
         & 1999.587 & 15.4 & 64 &  5 & 1.2 \\
         & 1999.587 & 43.2 & 64 &  7 & 1.5 \\
1031+567 & 1995.411 & 15.4 & 16 & 13 & 1.4 \\
         & 1999.587 & 15.4 & 64 &  5 & 1.0 \\
2352+495 & 1994.971 & 15.4 & 16 & 26 & 3.5 \\
         & 1994.971 & 43.2 & 64 &  6 & 1.6 \\
         & 1999.587 & 15.4 & 64 &  5 & 1.2 \\
\hline
\end{tabular}
\end{center}
$^a$ Total bandwidth refers to the sum of RCP and LCP bandwidths when
both were observed in 1994.971.

\clearpage

\begin{center}

TABLE 2 \\
\smallskip
S{\sc ource} I{\sc dentifications}
\smallskip

\begin{tabular}{l c c r r l}
\hline
\hline
Source & R.A. & Declination & $V$ & ID & \hfil $z$ \hfil \\
 (1)  & (2) & (3) & (4) & (5) & (6)   \\
\hline
\noalign{\vskip2pt}
0108+388 & 01 11 37.3118 & 39 06 28.110 & 22 & G & 0.6703 \\     
0710+439 & 07 13 38.1766 & 43 49 17.005 & 21 & G & 0.518 \\     
1031+567 & 10 35 07.0451 & 56 28 46.697 & 20 & G & 0.4597 \\    
2352+495 & 23 55 09.4539 & 49 50 08.362 & 20 & G & 0.237 \\     
\hline
\end{tabular}
\end{center}
Notes -- Col.(1): B1950 Source name according to IAU
convention. Cols.(2)-(3):J2000.0 right ascension and declination
(Patnaik \etal 1992).  Col.(4): Visual magnitude (Pearson \& Readhead 1988). 
Col.(5): Optical identification: G,
galaxy. Col.(6): Redshift (Lawrence \etal 1996).
\smallskip

\clearpage
\begin{center}

TABLE 3 \\
\smallskip
C{\sc ore} P{\sc roperties}
\smallskip

\begin{tabular}{l r r r r r}
\hline
\hline
Source &  $S_{\rm c}$(8 GHz) & $S_{\rm c}$(15 GHz) & $S_{\rm c}$(43 GHz) & $\alpha_{\rm c}$(8-15) & $\alpha_{\rm c}$(15-43)  \\
 (1)  & (2) & (3) & (4) & (5) & (6)   \\
\hline
\noalign{\vskip2pt}
0108+388 &  6.3 $\pm$ 1 & 14 $\pm$ 2 & $<$40 & 1.3 $\pm$ 0.4  & $<$1.0 \\
0710+439 & 42 $\pm$ 4 & 45 $\pm$ 5 & 37 $\pm$ 7 & 0.1 $\pm$ 0.3 & $-0.2$ $\pm$ 0.3 \\
1031+567 & $<$3  & $<2$ & -- & --  & -- \\
2352+495 & 13 $\pm$ 3 & 12 $\pm$ 3 & 6.5 $\pm$ 1 & $-$0.1 $\pm$ 0.7 & $-0.6$ $\pm$ 0.4 \\
\hline
\end{tabular}
\end{center}

\noindent
Notes -- Col.(1): Source name. 
Col.(2): Flux density of the core in mJy at 8.4 GHz from 
Xu (1994).
Col.(3): Flux density of core in mJy at 15 GHz. 
Col.(4): Flux density of core in mJy at 43 GHz. 
Col.(5): Spectral index of core at between 8.4 and 15 GHz.
Col.(6): Spectral index of core at between 15 and 43 GHz.
\smallskip

\clearpage

\begin{center}

TABLE 4 \\
\smallskip
G{\sc aussian} M{\sc odel} {\sc and} R{\sc elative} P{\sc roper} M{\sc otions for 0108+388}
\smallskip

\begin{tabular}{l r r r r r r r r r r}
\hline
\hline
Component & Epoch & \multicolumn{1}{c}{$S$} & \multicolumn{1}{c}{$r$} & \multicolumn{1}{c}{$\theta$} &\multicolumn{1}{c}{$a$} &\multicolumn{1}{c}{$b/a$}& \multicolumn{1}{c}{$\Phi$} & \multicolumn{1}{c}{$\mu$} & \multicolumn{1}{c}{$v$} & p.a. \\
       &  &  (Jy) &    (mas)   &  \multicolumn{1}{c}{(\dg)}&     (mas)  &   & \multicolumn{1}{c}{(\dg)}  & ($\mu$as/yr) & \multicolumn{1}{c}{($h^{-1}$ c)} & (\dg)  \\
\hline
\noalign{\vskip2pt}
C1\ldots & 1994.971 & 0.119 &  0.0     &      0.0  &  0.53 &  0.74 &  $-$30.5  \\
         & 1997.092 & 0.115 &  0.0     &      0.0  &  0.53 &  0.74 &  $-$30.5  \\
         & 1999.587 & 0.105 &  0.0     &      0.0  &  0.53 &  0.74 &  $-$30.5 &  reference \\
C2\ldots & 1994.971 & 0.033 &  0.894   &     91.85 &  0.82 &  0.72 &     66.4 \\
         & 1997.092 & 0.027 &  0.931   &     92.42 &  0.82 &  0.72 &     66.4 \\
         & 1999.587 & 0.023 &  0.963   &     92.90 &  0.82 &  0.72 &     66.4 & 16 $\pm$ 20 & 0.35 $\pm$ 0.44 & 106 \\
C3\ldots & 1994.971 & 0.010 &  2.510   &     55.98 &  0.29 &  1.0  &     0.0 \\
         & 1997.092 & 0.014 &  2.507   &     55.87 &  0.29 &  1.0  &     0.0 \\
         & 1999.587 & 0.012 &  2.615   &     55.59 &  0.29 &  1.0  &    0.0 &  26 $\pm$ 16 & 0.57 $\pm$ 0.35 & 48 \\
C4\ldots & 1994.971 & 0.037 &  3.664   &     54.76 &  0.89 &  0.20 &    74.7 \\
         & 1997.092 & 0.030 &  3.750   &     54.58 &  0.89 &  0.20 &    74.7 \\
         & 1999.587 & 0.024 &  3.664   &     54.26 &  0.89 &  0.20 &    74.7 & 23 $\pm$ 20 & 0.51 $\pm$ 0.44 & 44 \\
C5\ldots & 1994.971 & 0.054 &  4.022   &     53.04 &  0.17 &  0.28 &  $-$76.6 \\
         & 1997.092 & 0.061 &  4.097   &     53.60 &  0.17 &  0.28 &  $-$76.6 \\
         & 1999.587 & 0.069 &  4.170   &     53.99 &  0.17 &  0.28 &  $-$76.6 & 36 $\pm$ 2 & 0.79 $\pm$ 0.04 & 79 \\
C6\ldots & 1994.971 & 0.104 &  4.856   &     57.68 &  0.93 &  0.37 &  $-$84.8 \\
         & 1997.092 & 0.089 &  4.859   &     57.68 &  0.93 &  0.37 &  $-$84.8 \\
         & 1997.092 & 0.083 &  4.844   &     57.44 &  0.93 &  0.37 &  $-$84.8 & 4.7 $\pm$ 6 & 0.10 $\pm$ 0.12 & $-$67 \\
C7\ldots & 1994.971 & 0.172 &  5.830   &     59.84 &  0.47 &  0.48 &  $-$79.3 \\
         & 1997.092 & 0.152 &  5.841   &     59.88 &  0.47 &  0.48 &  $-$79.3 \\
         & 1999.587 & 0.130 &  5.878   &     59.98 &  0.47 &  0.48 &  $-$79.3 & 11 $\pm$ 2 & 0.24 $\pm$ 0.04 & 76 \\
\hline
\end{tabular}
\end{center}

\smallskip

\begin{center}
{\sc Notes to Table 4}
\end{center}

\smallskip
\noindent
NOTE -- Parameters of each Gaussian component of the model brightness
distribution: $S$, flux density; $r, \, \theta$, polar coordinates of
the center of the component relative to an arbitrary origin, with polar
angle measured from north through east; $a, \, b$, major and minor
axes of the FWHM contour; $\Phi$, position angle of the major axis
measured from north through east; $\mu$, relative proper motion of 
the component; $v$, relative projected velocity in units of $h^{-1}$ c
($h$ = H$_0$/100 km s$^{-1}$ Mpc$^{-1}$),
along the given position angle (p.a.).
\smallskip

\clearpage

\begin{center}

TABLE 5 \\
\smallskip
G{\sc aussian} M{\sc odel} {\sc and} R{\sc elative} P{\sc roper} M{\sc otions for 0710+439}
\smallskip

\begin{tabular}{l r r r r r r r r r r}
\hline
\hline
Component & Epoch & \multicolumn{1}{c}{$S$} & \multicolumn{1}{c}{$r$} & \multicolumn{1}{c}{$\theta$} &\multicolumn{1}{c}{$a$} &\multicolumn{1}{c}{$b/a$}& \multicolumn{1}{c}{$\Phi$} & \multicolumn{1}{c}{$\mu$} & \multicolumn{1}{c}{$v$} & p.a. \\
       &  &  (Jy) &    (mas)   &  \multicolumn{1}{c}{(\dg)}&     (mas)  &   & \multicolumn{1}{c}{(\dg)}  & ($\mu$as/yr) & \multicolumn{1}{c}{($h^{-1}$ c)} & (\dg)  \\
\hline
\noalign{\vskip2pt}
C2\ldots & 1994.971 & 0.051 &  0.0     &      0.0  &  1.30 &  0.62 &  $-$7.3  \\
         & 1999.587 & 0.043 &  0.0     &      0.0  &  1.30 &  0.62 &  $-$7.3 &  reference \\
B5\ldots & 1994.971 & 0.042 &  12.941  &     2.17 &  0.56 &  0.33 &  $-$13.1 \\
         & 1999.587 & 0.041 &  12.974  &     2.29 &  0.56 &  0.33 &  $-$13.1 & 9.3 $\pm$ 9.7 & 0.17 $\pm$ 0.17 & 42 \\
B4\ldots & 1994.971 & 0.121 &  14.205  &     2.03 &  0.46 &  0.25  &    15.8 \\
         & 1999.587 & 0.107 &  14.184  &     2.02 &  0.46 &  0.25  &    15.8  & 4.5 $\pm$ 8.7 & 0.08 $\pm$ 0.16 & 171 \\
B3\ldots & 1994.971 & 0.116 &  15.355  &     2.13 &  1.15 &  0.31  &  $-$25.5  \\
         & 1999.587 & 0.063 &  15.670  &     1.56 &  1.15 &  0.31  &  $-$25.5 & 76.0 $\pm$ 8.7 & 1.36 $\pm$ 0.16 & $-$24 \\
B2\ldots & 1994.971 & 0.115 &  15.854  &     2.66 &  1.05 &  0.29  &    13.1  \\
         & 1999.587 & 0.126 &  15.826  &     2.80 &  1.05 &  0.29  &    13.1  & 10.4 $\pm$ 8.7 & 0.19 $\pm$ 0.16 & 129 \\
A3\ldots & 1994.971 & 0.126 &  23.743  &     0.43 &  2.11 &  0.60  & $-$68.7  \\
         & 1997.092 & 0.095 &  23.679  &     0.60 &  2.11 &  0.60  & $-$68.7  & 20.6 $\pm$ 11 & 0.37 $\pm$ 0.20 & 133 \\
A2\ldots & 1994.971 & 0.185 &  24.333  &     1.16 &  0.51 &  0.65  &    51.4   \\
         & 1999.587 & 0.189 &  24.468  &     1.17 &  0.51 &  0.65  &    51.4   & 29.2 $\pm$ 8.7 & 0.52 $\pm$ 0.16 & 3 \\
\hline
\end{tabular}
\end{center}

\smallskip

\begin{center}
{\sc Notes to Table 5}
\end{center}

\smallskip
\noindent
See notes to Table 4.
\smallskip

\clearpage

\begin{center}

TABLE 6 \\
\smallskip
G{\sc aussian} M{\sc odel} {\sc and} R{\sc elative}  P{\sc roper} M{\sc otions for 1031+567}
\smallskip

\begin{tabular}{l r r r r r r r r r r}
\hline
\hline
Component & Epoch & \multicolumn{1}{c}{$S$} & \multicolumn{1}{c}{$r$} & \multicolumn{1}{c}{$\theta$} &\multicolumn{1}{c}{$a$} &\multicolumn{1}{c}{$b/a$}& \multicolumn{1}{c}{$\Phi$} & \multicolumn{1}{c}{$\mu$} & \multicolumn{1}{c}{$v$} & p.a. \\
       &  &  (Jy) &    (mas)   &  \multicolumn{1}{c}{(\dg)}&     (mas)  &   & \multicolumn{1}{c}{(\dg)}  & ($\mu$as/yr) & \multicolumn{1}{c}{($h^{-1}$ c)} & (\dg)  \\
\hline
\noalign{\vskip2pt}
W1\ldots & 1995.411 & 0.078 &  0.0     &    0.0  &  0.69 &  0.71 &     13.2  \\
         & 1999.587 & 0.080 &  0.0     &    0.0  &  0.69 &  0.71 &     13.2  &  reference \\
W2\ldots & 1995.411 & 0.097 &  1.075   &   62.24 &  1.68 &  0.54 &     71.5 \\
         & 1999.587 & 0.099 &  1.113   &   59.31 &  1.68 &  0.54 &     71.5  & 16.3 $\pm$ 12 & 0.27 $\pm$ 0.20 & 5 \\
E2\ldots & 1995.411 & 0.085 &  32.057  &   49.64 &  2.42 &  0.56 &     44.0  \\
         & 1999.587 & 0.092 &  32.211  &   49.74 &  2.42 &  0.56 &     44.0    & 37.6 $\pm$ 8.4 & 0.61 $\pm$ 0.14 & 70 \\
E1\ldots & 1995.411 & 0.065 &  33.699  &   50.34 &  0.72 &  1.00 &     0.00   \\
         & 1999.587 & 0.064 &  33.748  &   50.26 &  0.72 &  1.00 &     0.00   & 14.6 $\pm$ 4.8 & 0.24 $\pm$ 0.08 & 6 \\
\hline
\end{tabular}
\end{center}

\smallskip

\begin{center}
{\sc Notes to Table 6}
\end{center}

\smallskip
\noindent
See notes to Table 4.
\smallskip

\clearpage

\begin{center}

TABLE 7 \\
\smallskip
G{\sc aussian} M{\sc odel} {\sc and} R{\sc elative}  P{\sc roper} M{\sc otions for 2352+495}
\smallskip

\begin{tabular}{l r r r r r r r r r r}
\hline
\hline
Component & Epoch & \multicolumn{1}{c}{$S$} & \multicolumn{1}{c}{$r$} & \multicolumn{1}{c}{$\theta$} &\multicolumn{1}{c}{$a$} &\multicolumn{1}{c}{$b/a$}& \multicolumn{1}{c}{$\Phi$} & \multicolumn{1}{c}{$\mu$} & \multicolumn{1}{c}{$v$} & p.a. \\
       &  &  (Jy) &    (mas)   &  \multicolumn{1}{c}{(\dg)}&     (mas)  &   & \multicolumn{1}{c}{(\dg)}  & ($\mu$as/yr) & \multicolumn{1}{c}{($h^{-1}$ c)} & (\dg)  \\
\hline
\noalign{\vskip2pt}
C\ldots  & 1994.971 & 0.024 &  0.0     &     0.0   &  2.06 &  0.61 &   $-$81.0  \\
         & 1999.587 & 0.019 &  0.0     &     0.0   &  2.06 &  0.61 &   $-$81.0  &  reference \\
D\ldots & 1994.971 & 0.013 & 25.237   & $-$20.29  &  0.64 &  0.17 &       4.1  \\
        & 1999.587 & 0.012 & 25.310   & $-$20.18  &  0.64 &  0.17 &       4.1  & 19.1 $\pm$ 32 & 0.18 $\pm$ 0.31 & 13 \\
B5\ldots & 1994.971 & 0.084 & 30.104   & $-$15.72  &  0.53 &  0.53 &       8.4  \\
         & 1999.587 & 0.048 & 30.433   & $-$15.42  &  0.53 &  0.53 &         8.4  & 79.1 $\pm$ 11 & 0.76 $\pm$ 0.11 & 10 \\
B4\ldots & 1994.971 & 0.156 & 30.913   & $-$16.30  &  0.60 &  0.80 &   $-$54.3  \\
         & 1999.587 & 0.117 & 31.025   & $-$16.43  &  0.60 &  0.80 &  $-$54.3   &  28.6 $\pm$ 11 & 0.27 $\pm$ 0.11 & $-$48 \\
B3\ldots & 1994.971 & 0.075 & 31.390   & $-$15.66  &  1.56 &  0.39 &       0.4   \\
         & 1999.587 & 0.101 & 31.227   & $-$15.74  &  1.56 &  0.39 &       0.4   & 36.6 $\pm$ 11 & 0.35 $\pm$ 0.11 & 179 \\
B2\ldots & 1994.971 & 0.120 & 33.047   & $-$15.38  &  0.94 &  0.63 &   $-$25.4  \\
         & 1997.092 & 0.064 & 33.229   & $-$15.36  &  0.94 &  0.63 &   $-$25.4 & 39.4 $\pm$ 11 & 0.38 $\pm$ 0.11 & $-$12 \\
B1\ldots & 1994.971 & 0.085 & 33.360   & $-$17.03  &  1.20 &  0.70 &    $-$5.8  \\
         & 1999.587 & 0.058 & 33.592   & $-$17.00  &  1.20 &  0.70 &    $-$5.8  & 50.5 $\pm$ 11 & 0.48 $\pm$ 0.11 & $-$13 \\
A\ldots  & 1994.971 & 0.055 & 49.243   & $-$17.47  &  1.20 &  0.69 &   $-$68.2  \\
         & 1999.587 & 0.042 & 49.392   & $-$17.44  &  1.20 &  0.69 &   $-$68.2 & 32.7 $\pm$ 11 & 0.31 $\pm$ 0.11 & $-$8 \\
\hline
\end{tabular}
\end{center}

\smallskip

\begin{center}
{\sc Notes to Table 7}
\end{center}

\smallskip
\noindent
See notes to Table 4.
\smallskip

\clearpage
\end{document}